\documentclass[11pt, executivepaper]{article}
\usepackage[utf8]{inputenc}
\usepackage[T1]{fontenc}
\usepackage{natbib}
\usepackage{amsmath}
\usepackage{color}
\usepackage{amsfonts}
\usepackage{graphicx}
\usepackage{enumitem}
\usepackage{geometry}
\usepackage{hyperref}
\hypersetup{colorlinks= true, allcolors=blue}
\setcitestyle{aysep={}}

\title{\textbf{On the possibility of a realist ontological commitment in quantum mechanics}}

\author{Andrea Oldofredi\thanks{Contact Information: Universit\'e de Lausanne, Section de Philosophie, 1015 Lausanne, Switzerland. E-mail: Andrea.Oldofredi@unil.ch} \and Michael Esfeld\thanks{Universit\'e de Lausanne, Section de Philosophie, 1015 Lausanne, Switzerland. E-mail: Michael-Andreas.Esfeld@unil.ch}}

\begin{document}

\maketitle 

\begin{abstract}
This paper reviews the structure of standard quantum mechanics, introducing the basics of the von Neumann-Dirac axiomatic formulation as well as the well-known Copenhagen interpretation. We review also the major conceptual difficulties arising from this theory, first and foremost, the well-known measurement problem. The main aim of this essay is to show the possibility to solve the conundrums affecting quantum mechanics via the methodology provided by the primitive ontology approach. Using Bohmian mechanics as an example, the paper argues for a realist attitude towards quantum theory. In the second place, it discusses the Quinean criterion for ontology and its limits when it comes to quantum physics, arguing that the primitive ontology programme should be considered as an improvement on Quine's method in determining the ontological commitments of a theory.
\vspace{6mm}

\vspace{4mm}

\center
\emph{To be submitted to Tr\'opos. Rivista di Ermeneutica e Critica Filosofica}

\end{abstract}
\clearpage

\tableofcontents
\vspace{4mm}

\section{Introduction}

In the context of naturalized metaphysics we are typically guided by the best scientific theories available to inform our ontology; for instance, one may be ontologically committed to the existence of genes, molecules and elementary particles, accepting the content of biological, chemical and physical theories. 
 
Notorioulsy, this received view on ontological commitment in analytic philosophy was proposed by \cite{Quine:1948aa}. In this seminal essay the author claimed that ontological questions must be answered by looking at the content of the most advanced scientific theories at our disposal, proposing the following method: 
\begin{enumerate}
   \item Select a set of statements considered true from the best scientific theories available; 
   \item Express these statements in the language of first-order predicate logic (regimentation procedure);
   \item Be ontologically committed to all and only the bound variables appearing in existentially quantified sentences which make them true. 
\end{enumerate}
With Quine's words:
\begin{quote}
A theory is committed to those and only those entities to which the bound variables of the theory must be capable of referring in order that the affirmations made in the theory be true. (\cite{Quine:1948aa}, p. 33)
\end{quote}
To take an elementary example, endorsing what our most advanced physical theories claim, we accept the existence of electrons and hence, we consider true the quantified sentence ``Electrons exist''. Its translation into the language of first-order logic $\exists{x}E(x)$ is useful to make clear what entities we are ontologically committed to: in order for this sentence to be true, there must be in the world a physical entity to which the bound variable $x$ refers which satisfies the predicate $E(x)$, thus accepting this sentence we are committed to the existence of (at least) one electron. 

Interestingly, the first step requires that, relatively to a specific domain, there is a decisional element to establish which is the best theory among possible competitors in order to choose the sentences to regiment. Alternatively stated, before the regimentation procedure we need to select a theory which we endorse as (approximately) true among a set of possible rival frameworks. Although the application of this method to classical physical theories\footnote{The expression ``classical physical theories'' refers to a set of theories covering an extensive spectrum of scales, from descriptions of microscopic gasses provided by classical statistical mechanics, to cosmological effects treated by Einstein's theory of gravitation. Broadly speaking, they apply up to the length/energy scales where quantum interferences cannot be neglected.} is quite unproblematic, since these frameworks are ontologically well-defined theories, in the quantum context, as we will explain, one needs to do some work not only in order to choose what is the best framework to apply it, but also to spell out its limits. 

Analyzing the structures of classical theories, though remarkably different from one another, it is possible to state that (i) these theories provide a clear ontology, specifying the fundamental entities describing matter in space -- typical examples are the particle ontology instantiated by Newtonian mechanics or the field ontology proposed in Maxwell's electromagnetic theory -- and that (ii) physical objects can be described individually by the maximal set of properties they instantiate, whose values are independent of the performance of observations and always well-defined.

Against this background, the physics concerned with molecules, atoms and sub-atomic particles described by Quantum Mechanics (QM) calls into question a classical world-view. Contrary to the classical case, in quantum theory physical objects cannot be specified in terms of a maximal set of properties, since their values are not determinate until a measurement of a particular magnitude is performed: isolated systems in QM are objects with \emph{indeterminate} properties. For instance, reconsidering the previous example, the sentence $\exists{x}E(x)$ commits us to the existence of quantum objects without well defined position and momentum in space, since Heisenberg's uncertainty relations forbid the possibility to simultaneously measure with arbitrary precision the values of these magnitudes, concluding that electrons (and all quantum particles) do not have both a well defined position and a definite momentum. Thus, given the truth conditions imposed by QM, the application of Quine's method would imply a commitment to a quite peculiar ontology. However, one may object that it is not necessary to expect the retention of a classical world-view in quantum contexts. This objection, although fair, cannot resist that the peculiarities of QM are derived from conceptual and technical difficulties affecting its mathematical structure, like the measurement problem or the presence of mathematically ambiguous notions appearing within its axioms. Hence, these conundrums prevent us from considering it as a coherent description of the physical phenomena existing at the quantum length-scales. 

These facts suggest that one should look for a quantum theory that is immune to the conceptual difficulties affecting its standard formulation. To overcome the problematic aspects of QM a significant number of interpretations and alternative formulations have been presented.\footnote{\cite{Jaeger:2009aa} and \cite{Lewis:2016aa} discuss several proposals in detail.} In this paper we will consider the Primitive Ontology (PO) approach, a philosophical perspective which tries to solve the quantum puzzles by providing a precise methodology to construct physical theories with a well-defined ontology -- that is, theories which specify the theoretical entities representing matter in space and how they behave. These frameworks recover by construction the explanatory scheme of classical physical theories providing descriptions and explanations of quantum phenomena and processes in terms of well-defined entities moving and interacting in space and time, therefore avoiding obscure ontological statements about the indefiniteness of quantum objects. Notable examples, among others, are the Ghirardi-Rimini-Weber spontaneous collapse theories, in particular the variants GRWm and GRWf implementing a matter density field or flash ontology respectively, Nelsonian mechanics or Bohmian Mechanics (BM). In this essay we will concentrate on the latter proposal to show a possible way to recover a clear ontological commitment in the quantum realm. We will argue that since BM provides better explanations and descriptions of physical objects and processes at the quantum length scales, there are arguments to prefer BM over QM, and thus to consider the former a better candidate to apply Quine's method in the quantum context. Nevertheless, we will also discuss the general validity of this method and its limitations in the context of quantum physics, arguing that the PO perspective improves on this method to establish one's ontological commitments.
\vspace{2mm}

\section{Ontological Commitments in Standard Quantum Mechanics}

Although the appearance of the quantum hypothesis dates back to the first decade of the XXth century, with Planck's heuristic solution to the black-body radiation and Einstein's subsequent application to explain observational data concerning the photoelectric effect, the first coherent formulations of QM were proposed only in the mid-Twenties. On the one hand, Heisenberg, Born and Jordan developed matrix mechanics in their \emph{Drei-M\"{a}nner-Arbeit}, an algebraic formulation of quantum theory where physical magnitudes are expressed by means of matrices evolving in time, on the other, starting from de Broglie's relation, Schrödinger developed his wave mechanics, where physical states are represented by wave functions evolving in space and time according to a diffusion equation.\footnote{For a detailed account of historical developments of QM the reader should refer to \cite{Jammer:1974aa} and \cite{Seth:2013aa}.} 

Although the mathematical structures and the starting assumptions of these representations are profoundly different, their physical equivalence has been proved soon after. However, for a rigorous and systematic presentation of QM one should have waited until the textbooks of Dirac (\emph{The Principles of Quantum Mechanics}, 1930) and von Neumann (\emph{Mathematische Grundlagen der Quantenmechanik}, 1932). Remarkably, the von Neumann-Dirac formulation of quantum theory is, with minimal modifications, currently referred to as the ``standard'' or ``textbook'' formulation of QM.\footnote{\cite{Bub:1997aa} labeled this interpretation the ``basic approach'' to QM. \cite{Landsman:2017aa}, p. 2, provides a concise discussion of the differences between Dirac's and von Neumann's approach.} In this paper we stick to this interpretation to introduce the basic principles of the theory.

\subsection{The von Neumann-Dirac Principles of Quantum Mechanics}

According to this formulation of QM, also known as Princeton interpretation (see \cite{Jaeger:2009aa}, p. 117), a quantum system is described by a state vector $|\psi\rangle$, element of a complex vector space called Hilbert space $\mathcal{H}$, providing a \emph{complete} specification of its properties. In QM only \emph{operationally accessible} properties are considered magnitudes of quantum systems; they are represented by positive, linear Hermitian operators $\mathcal{A}$, acting on $\mathcal{H}$. Given a measurable quantity $A$, its possible values are the eigenvalues (real numbers) of the associated operator $\mathcal{A}$, whereas possible states in which a system may be found after a measurement of $A$ are represented by the eigenvectors of $\mathcal{A}$.
Implicitly we have defined the eigenvalue-eigenstate link, a core tenet of the von Neumann-Dirac formulation. With Dirac's words:
\begin{quote}
The expression that an observable `has a particular value' for a particular state is permissible in quantum mechanics in the special case when a measurement of the observable is certain to lead to the particular value, so that the state is in an eigenstate of the observable $[\dots]$. In the general case we cannot speak of an observable having a value for a particular state, but we can speak of its having an average value for the state. We can go further and speak of the probability of its having any specified value for the state, meaning the probability of this specified value being obtained when one makes a measurement of the observable. (\cite{Dirac:1947aa}, p. 253)
\end{quote}
As correctly pointed out by Jaeger, this quote eloquently characterizes the idea that quantum systems in isolation do \emph{not} instantiate properties with definite values; consequently, quantum mechanical objects possess indefinite properties prior experimental observation, a significant ontological difference w.r.t. classical objects. Unlike the classical case, (i) properties of quantum systems are not independent of observations, which become central in the microphysical context, and (ii) the theory provides only \emph{probabilities} for the possible measurement outcomes, as clearly expressed by the next postulate, the Born's rule. This axiom is crucial to connect the abstract Hilbert space formulation of QM with experimental results, and states that if the vectors $|\psi\rangle, |\phi\rangle$ represent two different physical states of a given system, there exists a probability $p(|\psi\rangle, |\phi\rangle)$ to find $|\psi\rangle$ in the state $|\phi\rangle$, given by the squared modulus of their scalar product on $\mathcal{H}$ $p(|\psi\rangle, |\phi\rangle)=|\langle\psi|\phi\rangle|^2$. 

The last two axioms of QM concern the dynamical evolution of physical systems. In the first place, the evolution in space and time of quantum systems is governed by a deterministic partial differential equation, the Schrödinger Equation\footnote{In SE $i$ is the imaginary unit, $\hbar=h/2\pi$ is the reduced Planck constant and $\hat{H}$ is the Hamiltonian operator representing the sum of kinetic and potential energy of the system.} (SE):
\begin{align}
\label{SE}
i\hbar\frac{\partial}{\partial{t}}|\psi\rangle =\hat{H}|\psi\rangle.
\end{align} 
SE has several properties among them linearity, implying that if two state vectors $|\psi_1\rangle,|\psi_2\rangle$ are both possible solutions of the same SE, then their linear combination (\emph{superposition}) $|\psi\rangle_s=\alpha|\psi_{1}\rangle+\beta|\psi_{2}\rangle$ is still a solution of the same SE.\footnote{Here $|\alpha|^2,|\beta|^2$, with $\alpha,\beta\in\mathbb{C}$, represent the probabilities to find the system in $|\psi_1\rangle,|\psi_2\rangle$ respectively. The normalization $|\alpha|^2+|\beta|^2=1$ means that with certainty we will find the system in one of the possible eigenstates.} Thus, the new superposed state $|\psi\rangle_s$ is also a consistent representation of the system. 

This fact is \emph{the} peculiarity of quantum mechanics, since in classical context \textcolor{blue}{$|\psi\rangle_s$} would represent a different state w.r.t. its superposed components. To better understand the effect of this shift, suppose to measure the $z$-spin of a quantum particle: the eigenstates in which it is possible to find it are only ``$z$-spin-up'' and ``$z$-spin-down'', but from linearity follows that also the superposition ``$z$-spin-up \emph{and} $z$-spin-down'' is a consistent description of the state in which the particle may be. This entails that prior to a spin measurement the particle has \emph{indefinite} spin, being neither in the $z$-spin-up state \emph{nor} in the $z$-spin-down state.\footnote{\cite{Hooker:1975aa} contains famous no-go theorems which prove the contextual nature of quantum theory.}

However, since in experimental observations such superpositions are never revealed, von Neumann and Dirac introduced the notion of projection of the state vector, which is distinctive of this formulation of QM. Suppose we measure a certain quantity represented by the operator $\mathcal{A}$ with eigenvalues $a_i$ and eigenvectors $|j\rangle$: if the system is described by $|\psi\rangle$, the Born's rule gives the probability to obtain a specific $a_j$ as measurement outcome, $p(a_j)=|\langle{j}|\psi\rangle|^2$. After the measurement's performance the system is projected into one of the possible eigenstates. Considering our previous example, the observation of the particle's $z$-spin (the interaction between the quantum particle and the experimental apparatus) causes a suppression of the SE and of the superposition with the consequent stochastic jump of the system in one of the possible spin eigenstates. These stochastic ``jumps'' make QM inherently probabilistic; Dirac viewed them as ``unavoidable disturbance'' of quantum systems in measurement situations:
\begin{quote}
When we measure a real dynamical variable $\xi$, belonging to the eigenvalue $\xi'$, the disturbance involved in the act of measurement causes a jump in the state of the dynamical system. From physical continuity, if we make a second measurement of the same dynamical variable immediately after the first, the result of the second measurement must be the same as the first. Thus after the first measurement has been made, there is no indeterminacy in the result of the second. Hence after the first measurement is made, the system is in an eigenstate of the dynamical variable $\xi$, [$\dots$]. In this way, we see that a measurement always causes the system to jump into an eigenstate of the dynamical variable that is being measured, the eigenvalue this eigenstate belongs to being equal to the result of the first measurement (\cite{Dirac:1947aa}, p. 36, reported in \cite{Jaeger:2009aa}, p. 120).
\end{quote}
Again, before the first measurement, the system's state is inherently indeterminate, since its properties depend strictly upon the act of observation. 
It is worth saying that the projection postulate has been introduced to reconcile the postulates of quantum theory with experimental evidence: assuming (i) that state vectors provide a complete description of quantum systems and (ii) that their motion would have been entirely described by the SE alone, the superposed states would have been amplified to the macroscopic scale. Thus, also macroscopic objects could have been in a superposition, contradicting the uniqueness and definiteness of measurement outcomes. This is the essence of the famous Measurement Problem (MP) of quantum theory, which can be stated more clearly as follows:\footnote{For a detailed discussion of the MP one may refer to \cite{Maudlin:1995aa} and \cite{Lewis:2016aa}. For a historical presentation of the quantum theory of measurement \cite{Jammer:1974aa} is still unsurpassed.}
\begin{enumerate}
   \item State vectors provide a complete description of quantum systems;
   \item State vectors evolve according to a linear dynamical equation (SE); 
   \item Measurements have a unique determinate outcome.
\end{enumerate}

Any pair of these statements is consistent and entails the falsity of the third one, but their conjunction generates inconsistencies with experimental evidence. Reconsidering the $z$-spin example, the device we use must have a ready state pointing in a neutral direction before the measurement, and two different pointers indicating different directions, let's say left ($L$) and right ($R$), representing the possible eigenstates in which we may find the system after the observation, $z$-spin up and $z$-spin down respectively. Feeding a particle in an eigenstate of $x$-spin into the classical device one obtains the following superposition in virtue of the SE:
\begin{align*}
\frac{1}{\sqrt{2}}(|x-up, L\rangle+|x-down, R\rangle).
\end{align*}

Thus, a microscopic superposition is amplified to macroscopic scale until a measurements is performed. However, since we never observe macroscopic superpositions, SE cannot provide a complete dynamical story for quantum objects. Experimental practice suggests that quantum systems are also subjected to stochastic jumps when they interact with measurement devices causing the instantaneously suppression of the unitary evolution. These interactions clearly have a macroscopic effect, namely the state of the experimental apparatus will be \emph{correlated} with the eigenstate (and relative eigenvalue) in which the quantum system will be found, so that we can actually observe a definite measurement result. This is the content of the projection postulate.\footnote{It should also be clear why quantum probabilities naturally emerge from the theory's formalism: if observations suppress SE leading to stochastic jumps, and if the state vectors provide complete descriptions of the quantum system, then the measurements' results must be inherently probabilistic, since they do not reveal pre-existing values of some property instantiated by quantum systems. Then, indeterminacy must be interpreted as an intrinsic feature of the quantum realm and should not be interpreted as ``lack of knowledge'' about initial conditions of quantum systems.}

Nonetheless, although the projection postulate makes the quantum formalism consistent with experimental evidence, it does not provide a good solution to the MP. In the first place it implies an indispensable, arbitrary and not precisely defined division between the observed system, which could also include the experimental apparatus, and the observer, who concludes every act of observation:\footnote{However, von Neumann explicitly denied the active role of the observer's consciousness in measurement situations: ``no matter how far we calculate [$\dots$] at some time we must say: and this is perceived by the observer. That is, we must always divide the world into two parts, the one being the observed system, the other the observer. In the former, we can follow up all physical processes (in principle at least) arbitrarily precisely. In the latter it is meaningless'' (\cite{vonNeumann:1955aa}, p 419). It is sufficient, then, to say that the observer's experience must be consistent with registered events in experimental situations. Indeed, even though observers play a crucial role in von Neumann's view, \cite{Whiteman:1971aa}, p. 71 notes that this interpretation is fully consistent with metaphysical realism: ``the classical tradition of simply located objects characterized independently of experiment, was presupposed by Born and von Neumann and imposed on the data with the help of an informal language of `particles' and `states' ''. Thus, although conceptually problematic, von Neumann's theory does not attribute to observers any creative role. Be that as it may, despite the heterogeneous personal beliefs on ontological matters there was no consensus among the fathers of the theory about its ontological content; see \cite{Guicciardini:2007aa} for historical details.} not only nothing in the formalism refers to the notion of observer, though it plays a fundamental role in the theory, but also the axioms of QM do not provide any description or explanation of the processes responsible for the suppression of the deterministic evolution provided by the SE. Thus, the notion of measurement, albeit pivotal within these axioms and taken as an unexplained primitive concept, is neither mathematically, nor physically well-defined. 

In sum, the projection postulates introduces an inconsistency between the dynamical laws governing the temporal evolution of quantum states, since it is not clear why and how measurement interactions cause the interruption of the continuous motion provided by the SE; in other words, the reasons which make ``being observed'' and ``not-being observed'' a relevant physical distinction for quantum systems remain entirely obscure. In addition to this, there is no explanation of what distinguishes a measurement interaction from a non-measurement interaction:
\begin{quote}
[w]hat the traditional theory did \emph{not} do is state, in clear physical terms, the conditions under which the non-linear evolution takes place. There were, of course, theorems that if one puts in collapses \emph{somewhere} between the microscopic and the macroscopic, then, for all practical purposes, it doesn't much matter \emph{where} they are put in. But if the linear evolution which governs the development of the fundamental object in one's physical theory occasionally \emph{breaks down} or \emph{suspends itself} in favor of a radically different evolution, then it is a physical question of the first order exactly under what circumstances, and in what way, the breakdown occurs. The traditional theory papered over this defect by describing the collapses in terms of imprecise notions such as ``observation'' or ``measurement'' (\cite{Maudlin:1995aa}, p. 9).
\end{quote}

Hence, having underlined these problematic aspects of QM, to apply Quine's method will lead to be ontologically committed to a peculiar world, where the nature of molecular, atomic and subatomic objects is not only inherently indeterminate, but this indeterminateness is moreover due to the problems related with the notion of measurement and observer. One should therefore look for a better theory to regiment.

\subsection{Princeton and Copenhagen}

Under many respects the von-Neumann-Dirac formulation of QM differs remarkably from the Copenhagen interpretation, which was mainly developed by Bohr and Heisenberg.\footnote{Interestingly, it is also controversial whether one may properly individuate a unitary set of accepted theses, given the significant divergences between the supporters of this `interpretation''. In this regard the reader may refer to \cite{Howard:2004aa} and \cite{Beller:1996aa}. For lack of space, here we concentrate only on a few aspects concerning the theory of measurement.} In the first place, Bohr never introduced explicitly the projection postulate, nor did he apply the quantum formalism to experimental devices being strongly convinced that one should describe them \emph{classically}, in virtue of the limited knowledge available of quantum systems expressed by the uncertainty relations; thus, quantum descriptions must be supplemented by classical concepts. Furthermore, not only the results of quantum measurements are necessarily expressed in terms of arrangements of macroscopic objects, but also the experimental procedure must be controllable and communicable in order to provide an objective description of quantum phenomena:\footnote{\cite{Landau:1981aa} in their monumental textbook on QM ironically recognized that classical mechanics should emerge as a limit of QM, but the latter needs the former for its foundations.}
\begin{quote}
it is decisive to recognize that, however far the phenomena transcend the scope of classical physical explanation, the account of all evidence must be expressed in classical terms. (\cite{Bohr:1958aa}, p. 39)
\end{quote}

Hence, measurement is a central notion in QM also according to Bohr's view, since properties of quantum objects strictly depend on the devices used in measurement situations: the changing experimental set-up will necessarily affect the nature of quantum systems. With Bohr's words:
\begin{quote}
The unambiguous account of proper quantum phenomena must, in principle, include a description of all relevant features of the experimental arrangement $[\dots]$. In the case of quantum phenomena, the unlimited divisibility of events implied in such an account is, in principle, excluded by the requirement to specify the experimental conditions. Indeed, the feature of wholeness typical of proper quantum phenomena finds its logical expression in the circumstance that any attempt at a well-defined subdivision would demand a change in the experimental arrangement incompatible with the definition of the phenomena under investigation. (\cite{Bohr:1963aa}, p. 3)
\end{quote}
 
Following this interpretation of QM, in agreement with the Princeton school, quantum objects have indeterminate properties in isolation. However, a central tenet of Bohr's view absent in the previous formulation of quantum theory is the \emph{complementarity principle}. In QM information concerning quantum systems is obtainable only through measurements, however, there are pieces of information that cannot be obtained simultaneously given the \emph{incompatibility of experimental protocols}, so that they cannot be represented by a unique quantum state of the examined system. Thus, the information obtainable by incompatible experiments is complementary. For instance, quantum objects have been claimed to have both corpuscular and wave descriptions, although it is not possible to show both these traits in a single observation, given the incompatibility of experimental procedures. To this regard, (\cite{Stapp:2009aa}, p. 113) claims that ``any preparation protocol that is maximally complete, in the sense that all the procedures are mutually compatible and are such that no further procedure can add any more information, can be represented by a quantum state, and that state represents in a mathematical form all the conceivable knowledge about the object that experiments can reveal to us''. Since quantum states represent the complete description of physical systems, it is clear that the nature of quantum objects essentially depends on experimental protocols and measuring devices and that observations do not reveal any pre-existing values of properties attributed to quantum systems. Interestingly, pairs of properties measurable with incompatible experimental protocols, such as momentum and position, cannot be revealed simultaneously, in agreement with Heisenberg's uncertainty relations. 

Instead the Princeton interpretation provides a different account of experimental situations: on the one hand, experimental protocols do not play a role in determining the properties of a quantum object, on the other, non-commuting observables cannot be measured simultaneously not having a set of common eigenstates. Thus, clearly, the von Neumann-Dirac interpretation provides a remarkably different explanation for the indeterminateness of quantum objects. Be that as it may, the theory remains silent about the ontological status of quantum objects in isolation:\footnote{\cite{Jaeger:2009aa} Chap. 3 and references therein provide a careful reconstruction of Bohr's philosophy.}

\begin{quote}
the quantum postulate implies that any observation of atomic phenomena will involve an interaction with an agency of observation not to be neglected. Accordingly, an independent reality in the ordinary physical sense can neither be ascribed to the phenomena nor to the agencies of observation. After all, the concept of observation is in so far arbitrary as it depends on which objects are included in the system to be observed. (\cite{Bohr:1934aa}, p. 3)
\end{quote}

As clearly expressed, also in this interpretation the division between classical and quantum regimes is as vague as the definition of the physical processes taking place during the interaction between quantum systems and classical devices. Furthermore, also in this case, the stochastic interruption of the SE in measurement situations does not find any precise description, nor explanation, so that it does not provide any satisfactory solution to the MP. Thus, also this interpretation of QM being based on the notion of measurement, inherits every conceptual problem presented in the previous paragraph. 
\vspace{2mm}

Taking into account a different Copenhagenist understanding of the quantum state, it is interesting to consider Heisenberg's subjective interpretation, which was mainly advanced in the Fifties.\footnote{The young Heisenberg was committed to a much more objective view of QM, being closer to Bohr ideas. In the Twenties he relegated the subjective element of quantum theory to the ignorance expressed by the probability function, which provides only a description of an ensemble of possible events, or potentialities, borrowing the Aristotelian terminology. These potentialities disappear in measurement situations. However, the subjective element in Heisenberg's view does not refer to the consciousness of the observer which is inactive in observational processes.} According to this account, the quantum state is a representation of the experimenter's \emph{knowledge} of a particular quantum system. The later Heisenberg said that:
\begin{quote}
When we are observing objects of our daily experience, the physical process transmitting the observation of course plays only a secondary role. However, for the smallest building blocks of matter every process of observation causes a major disturbance; it turns out that we can no longer talk of the behavior of the particle apart from the process of observation. In consequence, we are finally led to believe that the laws of nature which we formulate mathematically in quantum theory deal no longer with the particles themselves but with our knowledge of the elementary particles. The question whether these particles exist in space and time ``in themselves'' can thus no longer be posed in this form. We can only talk about the processes that occur when, through the interaction of the particle with some other physical system such as a measuring instrument, the behavior of the particle is to be disclosed (\cite{Heisenberg:1958aa}, pp. 99-100).
\end{quote}

This reading of the quantum state was also shared by other notable supporters of the Copenhagen interpretation, like Pauli or Peierls, who explicitly claimed that the statements of QM express fundamentally our knowledge of quantum systems.\footnote{This view contributed also to the development of Quantum Bayesianism, the most prominent contemporary subjective interpretation of QM.}

Accepting this interpretation of QM implies not only that a description of isolated quantum objects is unobtainable, but also that quantum mechanics is no longer a \emph{mechanical} theory, i.e. a framework providing descriptions and explanations of quantum phenomena in terms of objects moving and interacting in space and time. QM becomes an epistemological theory concerned the notion of knowledge of human observers implying a shift from an object-oriented ontology to a framework which describes the evolution (the updating) of agents' knowledge of some physical system. Therefore, the ontological commitment provided by a subjective reading of the wave function concerns agents' beliefs and remains silent on the ontological status of the physical systems these beliefs are about.  

In sum, if one applies Quine's method to these interpretations of QM either one is committed to the existence of objects with indefinite properties, or one is led to a subjective interpretation of QM that does not specify what kind of objects populate the quantum realm. 

\section{A Realist View in QM: the Theory of Local Beables}

In order to solve the conundrums afflicting the standard formulation of QM, (\cite{Bell:2004aa}, Chap. 7) proposed to focus the attention of physicists and philosophers working on the foundations of QM on the ontological problems deriving from its axioms. To formulate a set of principles not containing ill-defined notions, Bell's proposal was to spell out clearly, also in the quantum context, the ontology of a given physical theory $T$ via the specification of the set of theoretical entities with a reference to physical objects localized in space-time. Thus, he introduced the neologism \emph{beable}, from the English verb \emph{to be}, indicating what elements of a given physical theory $T$ refer to or represent real objects in the world and their properties, in opposition to what is only \emph{observ}-able, which plays a dominant role in the standard formulation of QM. To state it concisely, Bell's theory provides a specification of the primitive ontology of a given theoretical framework.\footnote{Here we consider the expressions ``primitive ontology'' and ``local beables'' synonyms since a precise assessment of their differences goes beyond the scope of the present paper.} 
\vspace{2mm}

Considering a physical theory $T$, its primitive ontology\footnote{The reader may refer to \cite{Allori:2013ab} and \cite{Esfeld:2014ac} for a more detailed explanation of this notion.} is a \emph{metaphysical} assumption defining the entities which cannot be further analyzed and in terms of more elementary notions.\footnote{The PO \emph{contextually depends} on the theoretical framework in which it is assumed. To this regard, \cite{Allori:2013ab}, p. 65 explicitly claims that ``there is no rule to determine the primitive ontology of a theory. Rather, it is just a matter of understanding how the theory was introduced, it has developed, and how its explanatory scheme works''.} These entities are the variables appearing in $T$'s equations with a \emph{direct} physical meaning, i.e. referring to (what according to $T$ are considered) \emph{real} objects precisely localized and moving in 3-dimensional physical space (or in 4-dimensional space-time). Furthermore, every physical phenomenon included in $T$'s domain is \emph{ontologically reduced} and explained via the dynamical evolution in space of these fundamental objects according to the particular laws of motion governing the behaviour of the PO, recovering in the quantum context the satisfactory and efficient explanatory scheme of classical physical theories. In this respect, we may also claim with Bell's words that
\begin{quote}
[t]he beables must include the settings of switches and knobs on experimental equipment, the currents in coils, and the reading of instruments. ``Observables'' must be \emph{made}, somehow, out of beables. The theory of local beables should contain, and give precise physical meaning to, the algebra of local observables (\cite{Bell:2004aa}, p. 52).
\end{quote}
This is to say that the primitive variables must \emph{connect} $T$ to our macroscopic ontology. It is worth noting that the theory of local beables follows the Bohrian idea for which ``the account of all evidence must be expressed in classical terms'': via the specification of the PO the vague expression ``classical terms'' acquires a precise meaning, since also the classical devices used to measure quantum properties are constituted by the beables of the theory, and are therefore treated in a mathematically and physically rigorous way as clearly stated in the above quotation. 

The central tenet of the PO approach to quantum physics is that every well-defined physical theory must satisfy the following requirements:
\begin{enumerate}
   \item A physical theory $T$ aims to provide a careful description of a specific domain of our world and to explain a specific set of phenomena;
   \item To provide this description, $T$ must implement a specific primitive ontology of objects moving in physical space (3-dimensional or 4-dimensional). These objects are fundamental in two senses: (i) they are not reducible to more elementary notions, (ii) the macroscopic objects of our ordinary experience must be reduced to the motion in space of the primitive variables;
   \item $T$ must provide a set of dynamical laws governing the motion of the PO;
   \item The mathematical structure of a given theory is naturally divided in two sub-structures: on the one hand, there are objets with a direct physical meaning, i.e. those entities referring to real objects in physical space, and on the other there are mathematical structures without a direct reference to physical objects. 
\end{enumerate}

These features constitute what \cite{Allori:2013ab} defines as \emph{the common structure} shared by the PO theories. Bohmian mechanics is an explicit example of such a theory. We will employ it to illustrate how the theory of local beables allows to solve the problems affecting the standard formulation of QM.

\subsection{An explicit Example: Bohmian Mechanics}

Bohmian mechanics is an alternative formulation of QM fully developed by the physicist David Bohm in two fundamental papers \cite{Bohm:1952aa} and \cite{Bohm:1952ab} and nowadays used by several theoretical physicists and quantum chemists.\footnote{The reader may refer to \cite{Bacciagaluppi:2009aa} for a historical reconstruction of the theory and to \cite{Freire:2015aa} for a sociological analysis of the elements involved in the abandonment of the Bohmian ideas. \cite{Durr:2009fk} is an excellent mathematical exposition of BM, whereas \cite{Oriols:2012aa} is concerned with some specific applications.} 

BM is a deterministic quantum theory of particles which move in three-dimensional physical space and follow continuous trajectories. Albeit this theory is statistically equivalent to the standard QM, their physical content is remarkably different since the former makes a precise metaphysical hypothesis concerning the intrinsic corpuscular nature of matter. In BM every physical fact is reduced to the motion of the Bohmian particles, which always have definite positions independently of any observation. 

According to this theory, physical systems are described by a couple $(\psi, Q)$, where the first element is the usual wave function and the second represents a specific $N$-particle configuration with positions $(Q_1,\dots, Q_N)$; these positions constitute the additional variables introduced by BM. The dynamics is composed by two laws of motion: on the one hand, the wave function $\psi$ evolves according to the usual SE \eqref{SE}, on the other, the motion of Bohmian particles is governed by the so-called \emph{guiding} equation:
\begin{align}
\label{vel}
\frac{dQ}{dt}=v^{\psi}_t(Q).
\end{align}

The vector velocity field on the right-hand-side in \eqref{vel} explicitly depends on the wave function, whose role is to guide the motion of the particles. The solutions of the guiding equation are integral curves corresponding to particles' trajectories. 

From \eqref{vel} it is easy to note that BM is a non-local theory, in perfect agreement with Bell's theorem. However, it is important to stress that there is no inconsistency with special relativity since Bohmian particles do not travel faster than light, thus, these non-local effects cannot be used to send signals at superluminal speed. Therefore, no \emph{operational} contradiction with relativity arises.

The last step we have to make in order to complete our brief presentation of BM is to guarantee the empirical, or statistical, equivalence w.r.t. the predictions of QM. The empirical equivalence is achieved via \emph{equivariance}: if at any arbitrary initial time $t_0$ the particle configuration is distributed according to $|\psi_{t_0}|^2$, then it will be so distributed for any later time $t$, preserving the Born's distribution (see \cite{Durr:2013aa}, Chapter 2 for the mathematical justification of this statement). 

The motivations to consider BM as a serious alternative to the standard quantum theory are very well known: not only the notorious measurement problem vanishes, but also BM does not rely on physically ill-defined notions such as \emph{measurement} and \emph{observer}, present instead in the axioms of standard QM. It is important to state that BM restores on the one hand an ontology of particles, and on the other an \emph{epistemic} interpretation of the quantum probabilities. The stochastic nature of the theory is a manifestation of our ignorance concerning the exact positions of the particles: according to BM, the maximal information at our disposal is always given by $|\psi|^2$, and randomness must be interpreted as a lack of knowledge absolutely detached from whatever sort of ontological indeterminacy of the quantum particles.\footnote{For details \cite{Durr:2013aa}, chapter 2, sec. 4-7. For a recent review of the literature on quantum probabilities in BM see \cite{Oldofredi:2016aa}.}
 
Having qualitatively introduced the bare bones of the theory, let us now discuss how it solves the MP. Consider a wave function which is in a superposition of two possible eigenstates of a two-valued operator (for instance the spin of a particle in one of the three possible directions):
\begin{align}
\psi=\alpha_1\psi_1+\alpha_2\psi_2 \nonumber
\end{align}
with $|\alpha_1|^2+|\alpha_2|^2=1$. Before the measurement we assume that the pointer of the experimental device is in the ready state $\Phi_0$, where the possible pointer positions will be $\Phi_1, \Phi_2$, indicating the possible eigenstates in which we may find the system after the observation. We label them position $1$ and $2$ for simplicity. 

According to quantum theory we know that, given the deterministic evolution provided by the SE, from the initial state in which the system under consideration and the apparatus are independent (and described by a product wave function), we obtain a macroscopic superposition:
\begin{align*}
\sum_{i=1,2}\alpha_i\psi_i\Phi_0\xrightarrow[evolution]{Schr\ddot{o}dinger}\sum_{i=1,2}\alpha_i\psi_i\Phi_i.
\end{align*}

The great merit of Bohmian mechanics comes from a simple idea: to describe quantum mechanically even the experimental apparatus, so that also macroscopic objects as the measurement devices are composed of Bohmian particles. This is in perfect agreement with Bell's theory of local beables introduced a few lines above.

In BM this experimental situation is described by a couple $(X_0,Y_0)$ where the former variable refers to the initial configuration of the particles of the system and the latter to the configuration of the particles that compose the apparatus at the initial time $t_0$. Then, given \eqref{SE} and \eqref{vel}, the total configuration of particles $(X_0,Y_0)$ evolves into another configuration $(X_t,Y_t)$ at time $t>t_0$ which is one of the possible eigenstates of the measured operator: the pointer will point to position 1 with probability $|\alpha_1|^2$ and to position 2 with $|\alpha_2|^2$. (For a detailed exposition see \cite{Durr:2009fk}, Chap. 9.) 

Considering the theory of measurement in BM, it would be immediately clear how measurement results of whatever observable depend strictly upon (i) the initial particle configuration of the system, which comprehends the observed system \emph{and} the experimental apparatus (which receives a quantum mechanical treatment in this context), and (ii) the dynamical equations of the theory, so that the final outcome is completely determined by the evolution in space and time of the total configuration of the particles involved (for details see Chap. 9 of \cite{Durr:2009fk} and \cite{Bohm:1952ab} sections 2, 3). Furthermore, contrary to the case of standard QM, in BM there is no collapse of the wave function: the dynamics of the system depends on the laws of motion governing the beables of the theory which prescribe a \emph{continuous} spatiotemporal evolution. Once one looks at the final configuration of particles, then one finds the pointer in one of the possible (macroscopic) positions corresponding to one of the possible outcomes. This means that within this theoretical framework nothing induces or causes stochastically the result of a given measurement.
Thus, Bohmian mechanics avoids by construction any reference to ill-defined notions: measurement outcomes receive an explanation in terms of the motion of the beables in space, becoming \emph{functions} of the primitive ontology. An explicit example in the context of BM is contained in \cite{Durr:2004c}, where the authors offer a carefully analysis of the reduction of spin measurements to position measurements, and consequently to the dynamical evolution of particles (for details see especially sec. 9). 

The case of BM is generalizable to every theory implementing a clear PO, since these frameworks provide a detailed theory of measurement to supply a rigorous description of the physical processes taking place in measurements situations, and therefore a detailed explanation of the obtained outcomes. This particular feature of the PO theories is crucial, since measurement constitutes the only connection between a given theoretical framework and experience: ``its analysis is therefore one of the most sensitive parts of any interpretation'' [of QM] (\cite{Jammer:1974aa}, p. 471). 

Ultimately, the macroscopic ontology of ordinary experience, to which experimental outcomes belong, is grounded in the primitive variables of a given theory and their motion in space and time, so that every physical fact within the theory's domain is reduced to them. In sum, we can agree with Bell in claiming that notions such as measurement, apparatus, observable and so on should be \emph{derived} from the primitive ontology. Observation is always theory-laden and a specific theory of measurement should be inferred from the fundamental structure of a theory, after all
\begin{quote}
[d]oes not any \emph{analysis} of measurement require concepts more \emph{fundamental} than the measurement? And should not the fundamental theory be about these more fundamental concepts? (\cite{Bell:2004aa}, p. 118)
\end{quote} 

\subsection{The status of the Wave Function in BM}

In contrast to the Copenhagen and the Princeton interpretations, BM offers an unambiguous formulation of quantum mechanics in which neither vague notions appear, nor the MP arises. Nonetheless, there is an open debate about its ontology (see e.g. \cite{Esfeld:2014ab}): if we apply the Quinean programme, the result is that we are committed to the existence of the particles \emph{and} the wave function, since the theory, if put in first order logic, quantifies over both of them. Indeed, according to Bohm himself and some contemporary supporters of his pilot-wave theory, like A. Valentini and P. Holland among others, the wave function is conceived as a field in physical space.\footnote{See \cite{Hubert:2017aa} for a recent defense of such a view about the wave function.} Hence, the sentence ``there exists an electron'', $\exists{x}Ex$, in this formulation of BM indicates that there exists a particle with well-defined values for position, mass and charge in physical space guided by a relative wave field, the wave function of the electron in question. 

However, for most interpreters this result is puzzling: the wave function is defined not on three-dimensional space, but on configuration space -- the mathematical space each point of which represents a possible configuration of particles in physical space (for $N$ particles, configuration space accordingly has 3$N$ dimensions). The puzzle then is how an object that exists in configuration space could influence the behaviour of objects in physical space. 

Since the debate about the status of the wave function is far from be solved, one may conclude that, once again, answering this ontological question requires genuine philosophical work instead of simply bringing these theories in the form of first order logic: before the regimentation of the sentences of BM, therefore, we need a precise answer to the metaphysical issue concerning the nature of the wave function. In this respect, the primitive ontology programme can be conceived as an improvement on the Quinean criterion for ontology. 

The PO, defined by the four criteria listed at the beginning of this section, is primitive also in the sense that the entities that it poses enter a physical theory $T$ as ultimate referents of $T$, which are not defined by their function for something else. In brief, the primitive ontology of a given theoretical framework is constituted by the set of those entities that, according to $T$, exist simply in the world.
According to this definition of primitiveness, the PO stands in contrast to the mathematical structures of a theory without a direct reference to physical objects; these latter ones are labeled \emph{non-primitive} variables in \cite{Allori:2008aa}, where the authors claim that the PO does not exhaust the entire ontology of a physical theory, since it comprehends also the structures in the $T$'s formalism whose function is to govern the behavior of the beables. Here, we follow a more recent literature so that we call these variables the \emph{dynamical structure} of $T$, which is made up by all those parameters that are introduced through their functional role for the evolution of those entities that constitute the primitive ontology. The wave function is the \emph{central} element of the dynamical structure of BM, because it is introduced in terms of its functional or causal role for the evolution of the particle configuration.

The dynamical structure is \emph{nomological} in the sense that it represents the behaviour of the elements of the primitive ontology, containing the parameters that are needed to formulate laws for the evolution of the beables. The distinction between primitive ontology and dynamical structure makes evident why the ontological commitment points towards the PO in the first place and why the ontological commitment to the dynamical structure is an open issue, as stressed a few lines above. 

In order to answer the question regarding the ontological status of the wave function and the other non-primitive variables, it is certainly possible to apply any one of the philosophical stances w.r.t. the natural laws to the dynamical structure as a whole: primitivism, according to which the dynamical structure is a further primitive over and above the primitive ontology (see e.g. \cite{Maudlin:2007aa} for that attitude with respect to the wave function); dispositionalism, a view in which the dynamical structure refers to dispositional properties of the physical objects in which the laws are anchored, think e.g. of gravitational mass in classical mechanics (dispositionalism with respect to the wave function in BM is set out in \cite{Belot:2012aa} and in \cite{Esfeld:2014ab}); Humeanism, where the dynamical structure is part of the best system -- that is, the system that achieves the best balance between simplicity and informational content in representing the evolution of the elements of the primitive ontology -- and hence does not call for an additional ontological commitment (that attitude is developed with respect to the wave function in \cite{Miller:2014aa}, \cite{Esfeld:2014aa}, \cite{Callender:2014aa}, \cite{Bhogal:2015aa}).

In a further step, one may then move on from the primitive ontology of a given theory to primitive ontology \emph{tout court}, that is, seek to work out a proposal about what the entities are that simply exist in the world, given our best physical theories. The rationale for doing so is that naturalistic metaphysics strives for an ontology of the natural world that is not relative to particular physical theories. One may even go as far as claiming that it is inappropriate to speak of the ontology of this or that theory. Ontology is about what there is. It goes without saying that our access to what there is comes through the representations that we conceive in terms of physical theories. But this does not necessarily imply that ontology is relative to particular theories. In other words, the idea for a fundamental ontology is to search for an answer to the following question: What is a minimal set of entities that form an ontology of the natural world, given our well-established physical theories? An accredited candidate for an answer to that question, both in classical and in quantum physics, is an ontology composed by point particles that are characterized only by their relative positions (that is, by their distances to each other) and the change in position (see \cite{Esfeld:2017aa} for making the case for that answer).

\section{Conclusion}

In this paper we have argued that the PO approach provides a sound methodology to construct theories avoiding the difficulties affecting the standard formulation of QM: according to this framework, the structure of a physical theory is well-defined if and only if its basic entities are specified and their equations of motion are given. In this manner, one is able to obtain an explanation of physical phenomena in terms of the evolution of the primitive ontology in space. 

From our discussion it emerged that Quine's method is not rich enough to distinguish between the mathematical structures provided with a direct physical meaning and those which are not; therefore, it cannot be \emph{alone} a secure guide to one's ontological commitment in microphysical contexts, not only given the pitfalls inherent to the quantum formalism, but also since the ontological status of the wave function is not definitively established, so that it still depends contextually on one's interpretation of $\psi$. It is clear, then, that one's ontological commitment depends on the answer given to this question. 
To this regard, however, the major methodological contribution conveyed by the PO is that the mathematical, physical and philosophical aspects of a particular theoretical framework should be clearly separated, so that it is always possible to interpret its elements (primitive and non-primitive) unambiguously. The main message of this paper, therefore, is to consider the PO programme as a useful guide in search for an ontology in the quantum realm to be added to Quine's method.

Furthermore, this perspective has been recently extended also to the realm of Quantum Field Theory (see \cite{Struyve:2010aa} and references therein for an overview as well as, most recently, \cite{Deckert:2016ac}) where, once again, Bohmian theories show the concrete possibility to consistently apply the PO methodology to recover QFT's phenomenology via the definition of ontologically clear alternative formulations. Although these theories may be still speculative or involve a partial reformulation of quantum field theory, they have the merit to show possible solutions to the conceptual issues present in the standard formulation, indicating a way to find a clear ontology also in the quantum theory of fields. 

\bibliographystyle{apalike}
\bibliography{PhDthesis}
\end{document}